\documentclass{desyproc}

\begin{document}

\title{Search for solar axions produced by Compton process and bremsstrahlung using
 the resonant absorption and axioelectric effect}

\author{{A.V.~Derbin, I.S.~Dratchnev, A.S.~Kayunov, V.N.~Muratova, D.A. Semenov, E.V. Unzhakov}\\ \\
        St.Petersburg Nuclear Physics Institute, Gatchina, Russia 188300}

\contribID{Derbin_Alexander}

\desyproc{DESY-PROC-2013-XX}
\acronym{Patras 2013} 
\doi  

\maketitle

\begin{abstract}
The search for resonant absorption of Compton and bremsstrahlung solar axions by $^{169}$Tm nuclei have been performed.
Such an absorption should lead to the excitation of low-lying nuclear energy level: $A+^{169}$Tm $\rightarrow
^{169}$Tm$^*$ $\rightarrow ^{169}$Tm $+ \gamma$ (8.41 keV). Additionally the axio-electric effect in silicon atoms is
sought. The axions are detected using a Si(Li) detectors placed in a low-background setup. As a result, a new model
independent restrictions on the axion-electron and the axion-nucleon coupling: $g_{Ae}\times|g^0_{AN}+ g^3_{AN}|\leq
2.1\times10^{-14}$ and the axion-electron coupling constant: $|g_{Ae}| \leq 2.2\times 10^{-10}$ has been obtained.  The
limits leads to the bounds $m_{A}\leq$ 7.9 eV and $m_{A}\leq$ 1.3 keV for the mass of the axion in the DFSZ and KSVZ
models, respectively (90\% C.L.).
\end{abstract}

\section{The axions spectra and the rate of axions resonant absorbtion by $^{169}\rm{Tm}$ nucleus}
\begin{figure}[hb]
\centerline{\includegraphics[width=0.5\textwidth]{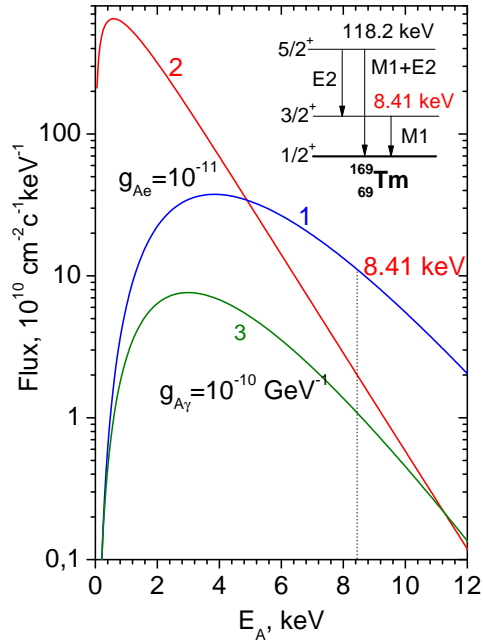}} \caption{1,2 - the spectra of the axions produced by the
Compton process and the bremsstrahlung, correspondingly ($g_{Ae}=10^{-11}$, $m_A=0$) \cite{Der11}. 3 - spectrum of the
axions produced by Primakoff effect ($g_{A\gamma}=10^{-10}\rm{GeV}^{-1}$). The level scheme of $^{169}$Tm nucleus is
shown in the inset \cite{Der11}.}\label{fig1.eps}
\end{figure}

If the axions or other axion-like pseudoscalar particles couple with electrons then they are emitted from Sun by the
Compton process and by bremsstrahlung \cite{Kat75}-\cite{Raf86}. The expected spectrum of axions was calculated using
theoretical predictions for the Compton cross section given in \cite{Pos08,Gon08} and the axion bremsstrahlung due to
electron-nucleus collisions given in \cite{Zhi79}. The axion flux is determined for radial distribution of the
temperature, density of electrons and nuclei given by BS05(OP) Standard Solar Model \cite{Bah05} based on high-Z
abundances \cite{Asp06}. The results of our calculations presented in Fig.\ref{fig1.eps} \cite{Der11}.

As a pseudoscalar particle, the axion should be subject to resonant absorption and emission in the nuclear transitions
of a magnetic type. In our experiment we have chosen the $^{169}$Tm nucleus as a target. The energy of the first
nuclear level (3/2$^+$) is equal to 8.41 keV, the total axion flux at this energy is $g_{Ae}^2\times1.34\times10^{33}
\rm{cm}^{-2}\rm{s}^{-1}\rm{keV}^{-1}$. The 8.41 keV nuclear level discharges through $M1$-type transition with
$E2$-transition admixture value of $\delta^2$=0.11\% and the relative probability of $\gamma$-ray emission is
$\eta=3.79\times10^{-3}$ \cite{Bag08}.

The cross-section for the resonant absorption of the axions with energy $E_A$ is given by the expression that is
similar to the one for $\gamma$-ray resonant absorption, but the ratio of the nuclear transition probability with the
emission of an axion $(\omega_{A})$ to the probability of magnetic type transition $(\omega_{\gamma})$ has to be taken
into account \cite{Der11}. The $\omega_A/\omega_\gamma$ ratio calculated in the long-wave approximation, depends on
isoscalar $g_{AN}^{0}$ and isovector $g_{AN}^{3}$ coupling constants and parameters depending on the particular nuclear
matrix elements \cite{Don78,Avi88,Hax91}.

As a result the rate of axion absorption by $^{169}$Tm nucleus dependent only on the coupling constants is (the
model-independent view) \cite{Der11}:
\begin{equation}\label{rategamag0g3}
    R_A=1.55\times10^5 g_{Ae}^2(g_{AN}^{0}+g_{AN}^{3})^2(p_A/p_\gamma)^3, \rm{s}^{-1}.
\end{equation}
Using the relations between $g_{AN}^0$, $g_{AN}^3$ and axion mass given by KSVZ model, the absorption rate can be
presented as a function of $g_{Ae}$ and axion mass $m_A$ ($m_A$ in eV units):
\begin{equation}\label{rategama}
    R_A=5.79\times10^{-10}g_{Ae}^2m_A^2(p_A/p_\gamma)^3,  \rm{s}^{-1}.
\end{equation}

\section{ Spectra of massive solar axions and a cross section for the axioelectric effect}
If the mass of the axion is several keV, the expected solar axion spectra changes significantly and depends on the
particular $m_{A}$ value \cite{Der12}. To calculate spectra we used the procedure similar described above. The spectra
of solar axions were determined for different values of $m_{A}$ \cite{Der12}.

An axion interacting with an electron should undergo axio-electric absorption, which is an analog of the photoelectric
effect. Silicon atoms entering into the composition of a Si(Li) detector were used in our experiment as targets for the
axio-electric effect. The cross section of the axioelectric absorption was calculated by the formula
\begin{equation}
\sigma_{abs}(E_{A})=\sigma_{pe}(E_{A})\frac{g^{2}_{Ae}}{\beta}\frac{3E^{2}_{A}}{16\pi\alpha
m_e^2}\left(1-\frac{\beta}{3}\right)
\end{equation}
where $\sigma_{pe}$ is the cross section for the photoelectric effect and $\beta= v/c = p_{A}/E_{A}$ is the velocity of
the axion. At $\beta \rightarrow $ 1 and $\beta \rightarrow $ 0, this formula coincides with the cross sections for
relativistic and nonrelativistic axions obtained in \cite{Pospelov:2008, Derevianko:2010} and provides an extrapolation
approximately linear in $\beta $, which ensures a sufficient accuracy for the case under consideration.

\begin{figure}[hb]
\centerline{\includegraphics[width=0.5\textwidth]{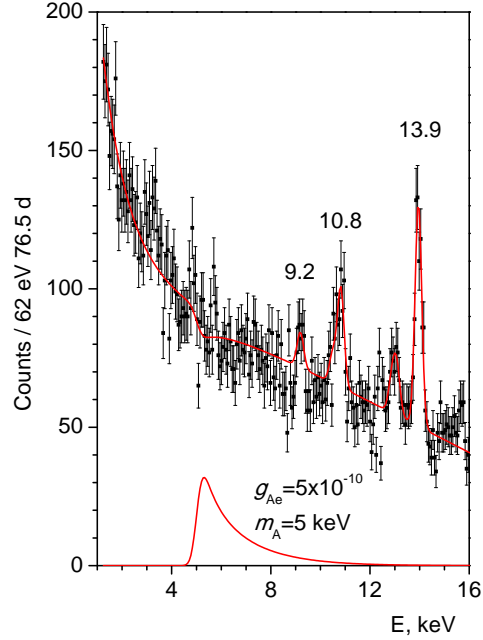}} \caption{Spectrum of the signals of the Si(Li) detector in
the range of (1--16) keV with the optimal fit for $m_{A} = 5$ keV. The energies of the Gaussian peaks are given near
them. The expected spectrum is shown for the case of the detection of axions with $m_{A} = 5$ keV and $g_{Ae} = 5
\times 10^{-10}$.}\label{fig2.eps}
\end{figure}

The expected signal from 5 keV axions undergouing axioelctric effect in Si-detector is shown in Fig.\ref{fig2.eps}.

\section{Experimental setups}
To search for quanta with an energy of 8.41keV due to axion resonant absorption, the planar Si(Li) detector with a
sensitive area diameter of 66 mm and a thickness of 5 mm was used. The detector was mounted on 5 cm thick copper plate
that protected the detector from the external radioactivity. The detector and the holder were placed in a vacuum
cryostat and cooled to liquid nitrogen temperatures. A $\rm{Tm}_2\rm{O}_3$ target of 2 g mass was uniformly deposited
on a plexiglas substrate 70 mm in diameter at a distance of 1.5 mm from the detector surface. External passive
shielding composed of copper, iron and lead layers was adjusted to the cryostat and eliminated external radioactivity
background by a factor of about 500.

To search for axioelectric effect we used a Si(Li) detector with a sensitive-region diameter of 17 mm and a thickness
of 2.5 mm. The detector was placed in a vacuum cryostat with the input beryllium window 20 $\mu$m thick. The window was
used for energy calibration and determination of the detection efficiency of gamma-ray photons in order to find the
sensitive volume of the detector. The detector was surrounded by 12.5 cm of copper and 2.5 cm of lead, which reduced
the background  of the detector at an energy of 14 keV by a factor of 110 as compared to the unshielded detector.

The experimental setups were located on the ground surface. Events produced by cosmic rays and fast neutrons were
registered by an active shielding consisting of five plastic scintillators $50\times50\times12$ cm in size. The rate of
50 $\mu$s veto signals was 600 counts/s, that lead to $\approx$ 3\% dead time. More details of experiments one can find
in \cite{Der11,Der12}.

\begin{figure}[hb]
\centerline{\includegraphics[width=0.5\textwidth]{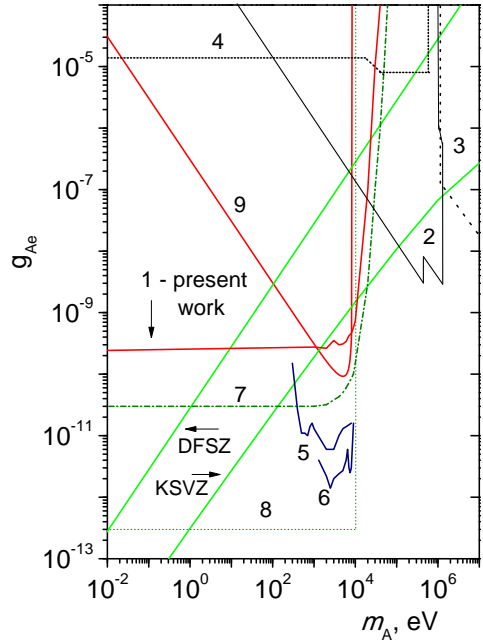}} \caption{Bounds for the axion-electron coupling constant:
(1) Si-axioelectrical effect \cite{Der12}, (2) reactor experiments and solar axions with energies of 0.478 and 5.5
MeV,(3) beam dump experiments, (4) decay of orthopositronium, (5) CoGeNT, (6) CDMS, (7) bound for the axion luminosity
of the Sun, (8) red giants and (9) experiment with $^{169}\rm{Tm}$. The regions of excluded values lie above the
corresponding lines. The inclined lines show the $g_{Ae}$ values in the DFSZ and KSVZ ($E/N=8/3$)
models.}\label{fig3.eps}
\end{figure}

\section{Results}
The upper limit on axions absorption rate by $^{169}\rm{Tm}$ nucleus $R_A \leq 5.43\times10^{-23}  \rm{s}^{-1}$ set by
our experiment limits the possible values of coupling constants $g_{Ae}$, $g_{AN}$ and axion mass $m_A$. According to
(\ref{rategamag0g3}) and (\ref{rategama}) and taking into account the approximate equality of the axion and
$\gamma$-quantum momenta $(p_A/p_{\gamma})^3 \simeq 1$ for $m_A\leq2$ keV we obtain (at 90\% c.l.):
\begin{equation}\label{lim1}
g_{Ae}\times |(g_{AN}^0+g_{AN}^3)| \leq 2.1\times 10^{-14}
\end{equation}
\begin{equation}\label{lim2}
g_{Ae}\times m_A \leq 3.1\times 10^{-7}\;\rm{eV}
\end{equation}
The restriction (\ref{lim1}) is a model independent one on axion (or any other pseudoscalar particle)  couplings with
electron and nucleons. The result (\ref{lim2}) presented as a restriction on the range of possible values of $g_{Ae}$
and $m_A$ (KSVZ relations between $g_{AN}$ and $m_A$ are used) allows one to compare our result (Fig.\ref{fig3.eps},
line 9) with results of other experiments restricting $g_{Ae}$  (Fig.\ref{fig3.eps}). The limits on $g_{Ae}\times m_A$
for DFSZ axion lie in the range ($0.33-1.32$) of the restriction. As we mention in \cite{Derbin:2009,Der11} the
sensitivity of experiment with $^{169}\rm{Tm}$ can be increased significantly (in $\sim 10^6$ times) by introducing the
Tm target inside the sensitive volume of detector having high energy resolution, at present, cryogenic detectors have
the best option.

When looking for the axioelectric effect the measured spectrum is fitted by the sum of an exponential function,
describing the smooth background and the response function for axions $S(E, m_{A})$:
\begin{equation}
N(E) = a+b\exp(cE)+g^{4}_{Ae}S(E,m_{A})N_{Si}T\label{fit_a_e}
\end{equation}
Here, $N_{Si}$ is the number of silicon atoms in the sensitive volume of the detector and $T = 6.61 \times 10^{6}$ is
the live time of measurement.  The upper bound for $|g_{Ae}|$ at $m_A$=0 is
\begin{equation}
|g_{Ae}|\leq 2.2 \times  10^{-10}\label{mainlimit}
\end{equation}
at 90\% C.L. Limit (\ref{mainlimit}) is a model independent bound for the coupling constant of the axion or any other
pseudoscalar relativistic particle with the electron.

For nonrelativistic axions the fitting range was expanded to 16 keV. To describe the experimental spectrum in a wide
range, function \ref{fit_a_e} was supplemented by a linear term for describing the continuous background and six
Gaussians for describing the peaks of the characteristic Np X rays manifested in the measurements
\cite{Derbin:2009,Derbin:2009A}. The fit results and $S(E, m_{A})$ for $m_{A}$ = 5 keV are shown in Fig.\ref{fig2.eps}.

The maximum deviation of $g^{4}_{Ae}$ from zero for all tested (from 1 to 10 keV with a step of 1 keV) $m_{A}$ values
is 2.5$\sigma$. The upper bounds obtained for $|g_{Ae}|$ at various $m_{A}$ values are shown in Fig.\ref{fig3.eps}
(line 1) in comparison with the other experimental results.

\section{Acknowledgments}

This work was supported by RFBR grants 13-02-01199-a and 13-02-12140-ofi-m.



\begin{footnotesize}

\end{footnotesize}


\end{document}